\documentclass{article}
\usepackage[a4paper, total={6in, 8in}]{geometry}
\usepackage[export]{adjustbox}
\usepackage[english]{babel}
\usepackage[utf8]{inputenc}
\usepackage{capt-of}
\usepackage{amsmath}
\usepackage{mathtools}
\usepackage{amsmath}
\usepackage{mathtools}
\usepackage{adjustbox}
\usepackage{array}
\usepackage{booktabs}
\usepackage{dsfont}
\usepackage{multirow}
\usepackage{hhline}
\usepackage[utf8]{inputenc} 
\usepackage[T1]{fontenc}    

\usepackage[hyphens]{url}
\usepackage[hidelinks]{hyperref}  
\usepackage{amsfonts}       
\usepackage{amsthm}
\usepackage{float}
\usepackage{graphicx}
\usepackage{xcolor}
\usepackage{amsthm,mathtools}
\usepackage[skip=10pt plus1pt, indent=40pt]{parskip}

\title{The Fairness of Machine Learning in Insurance: \\New Rags for an Old Man?}
\author{Laurence Barry$^{a}$ \& Arthur Charpentier$^{b}$\\
        \small $^{a}$ Chaire PARI, Fondation Institut Europlace de Finance, Paris \\
        \small $^{b}$ Université du Québec à Montréal (UQAM), Montréal (Québec), Canada \\}
\date{May 2022}

\begin{document}
\maketitle

\begin{abstract}
Since the beginning of their history, insurers have been known to use data to classify and price risks. As such, they were confronted early on with the problem of fairness and discrimination associated with data. This issue is becoming increasingly important with access to more granular and behavioural data, and is evolving to reflect current technologies and societal concerns. By looking into earlier debates on discrimination, we show that some algorithmic biases are a renewed version of older ones, while others show a reversal of the previous order. Paradoxically, while the insurance practice has not deeply changed nor are most of these biases new, the machine learning era still deeply shakes the conception of insurance fairness.
\end{abstract}

\vspace{1cm}

\begin{flushright}
``{\em Technology is neither good nor bad; nor is it neutral}'' \cite{kranzberg1986technology}
\end{flushright}

\section{Introduction}

Since the beginning of their history, insurers have been quantifying human phenomena, collecting and using data to classify and price risks. This practice is not trivial: insurance plays a major role in industrialized societies in opening or closing life opportunities (\cite{horan2021insurance}; \cite{baker2002embracing}). As such, insurers were confronted early on with the equity issues associated with data. As early as 1909, the Kansas insurance regulator thus drew the contours of fair rating practice: he defined a rating as "not unfairly discriminatory" if it treats similar people in the same way (\cite{Frezal_Barry_2019},  \cite{miller2009disparate}). 

Highlighting a bias means taking a critical stance on a calculation and shedding light on its political implications. This contestation is also culturally and historically situated. Based on this principle, certain parameters in insurance pricing were challenged during the 20th century, hence refining a typology of potential biases linked to classical data processing. The first part presents a historical perspective on insurance biases, showing how they were reformulated, clarified and reorganized with cultural and technological developments.

The second part describes the changes implied by the recent emergence of big data and new algorithms on insurance, and their induced biases. These debates actually revive and renew, with obvious points of continuity and rupture, older debates linked to discrimination in insurance. Conceptually however, these changes are shaking up insurance practice: the 'individualization' of risks, rendered possible by these new technologies, is indeed now considered as fairer than their pooling.  Finally, the last part discusses the ethical issues of big data for insurance, in comparison with various notions of algorithmic fairness. 

\section{Insurance and fairness: retrieving types of bias from historical debates}

\subsection{Pricing Practices Before Machine Learning}

Insurance consists in the pooling of uncertainty: the contribution of the many makes it possible to compensate for the accidents of the few unluckiest. In its crudest form, the insurance risk premium is the mathematical expectation, on the group at stake, of the cost of the accident. Without competition between insurers, a single rate for all, as the average risk for the whole population, would do. Competition leads however to the threat of anti-selection: by segmenting the premium, insurer A can attract the best risks, thereby increasing his share at the expense of his competitors, who will make losses if they do not adopt a similar strategy. Segmentation therefore quickly becomes the rule of the game. 

For a very long time, this segmentation consisted in the creation of supposedly homogeneous classes, on which the risk was estimated on average (\cite{charpentier103}). Before any calculation, the actuary's had to choose variables, a choice that projected homogeneity on the world, in two main ways: in the choice of what to ignore on the one hand (as what is not collected contains some heterogeneity that will not be seen); in the categorization of what is collected on the other hand, which again leads to crushing potential differences. 
Gradually, the idea of a ``perfect rate" emerged, in which the class would only include truly identical risks. Following \cite{charpentierMASSNV2}, and admitting that $\Theta$ is the variable that would perfectly characterize the risk, we obtain the decomposition of Table \ref{tab:risk:share},
$$
\text{Var}[Y]=\underbrace{\mathbb{E}\big[\text{Var}[Y|\Theta]\big]}_{\to\text{ insurer}}
+\underbrace{\text{Var}\big[\mathbb{E}[Y|\Theta]\big]}_{\to\text{ policyholder}}.
$$

\begin{table}[!ht]
    \centering
\begin{tabular}{|l|cc|}
\hline
 & Policyholder & Insurer\\
 \hline
Loss & $\mathbb{E}[Y|\Theta]$ & $Y-\mathbb{E}[Y|\Theta]$ \\
Average loss & $\mathbb{E}[Y]$ & 0\\
Variance & $\text{Var}\big[\mathbb{E}[Y|\Theta]\big]$ & $\text{Var}\big[Y-\mathbb{E}[Y|\Theta]\big]$\\
\hline
\end{tabular}
    \caption{The sharing of losses and variance}
    \label{tab:risk:share}
\end{table}

The variance on the portfolio is thus distributed between the policyholders who pay premiums proportional to their risk (captured by theta) and the insurer who carries the residual variance, unexplained by theta. In the 1980s, \cite{de1984rate} claimed that the increasing data collection and computational capabilities made it possible to refine the explained variance (and segmented premiums), thereby reducing the portion carried by the insurer. 

However, the theta parameter, supposedly perfectly catching the risk, is never known. This uncertainty is the very basis of insurance: in a model for death benefits for instance, the probability of death can be estimated more precisely (some people have a 1 in 10,000 chance of dying and others a 1 in 1,000 chance), but it is impossible to predict who will die in a year. This fundamental residual uncertainty will irreducibly remain the insurer’s responsibility, thus covered by the law of large numbers. Classification is then rethought as a means of approaching theta: the goal is no longer to simply counter anti-selection with ever finer classes, but also to have the unexplained variance converge towards a minimum. The actuarial practice does not change, even if its meaning does.

It is in this adjustment framework that the notion of bias appears: in the hypothesis that an exact calculation of the risk is possible, it occurs if the classification is imperfect and leads to the mispricing of certain groups, or persons, creating cross-subsidies between insureds (\cite{walters1981risk}; \cite{de1996segmentering}).

\subsection{Historical Controversies on Insurance Pricing}

From the 1960s onwards in the United States, the classification of risks was called into question in two main respects. First, in the context of the struggle for black civil rights, the practice of ``red lining" or the exclusion of certain geographical areas from insured portfolios was criticized. Later, at the end of the 1970s, feminist movements tried to counter the use of gender in underwriting and pricing (\cite{horan2021insurance}). A detailed examination of the arguments allows to specify different aspects of what can be called ``biased pricing", leading to a typology of pre-machine learning insurance biases that will serve our examination of machine learning in the next section.

One way to look at classification is to take it as a method for an ex-ante distribution of future costs, always more or less arbitrary. As such, the quantification work (of the statistician or actuary) is criticized because it is fed by an always socially constructed vision of the world  (\cite{desrosieres2008argument}). \cite{glenn2000shifting} points out that, like the Roman god Janus, an insurer's risk selection process has actually two faces: on the one side there is the face of numbers, actuarial tables, and statistics, that claims objectivity and rationality. But on the other side, there is the face of stories and subjective judgement. For Glenn, the actuary creates a myth in which decisions appear to be objective when in fact they are based on a great deal of subjectivity, prejudice, and stereotyping. These stereotypes come upstream the construction of the actuarial tables, in the stories that insurance technicians (actuaries and underwriters) tell each other, which lead them to collect one variable or another. Indeed, as data collection is still done in the form of questionnaires, it is necessarily oriented by the pre-conception of risk held by the designer, itself socially constructed. These prejudices also appear downstream, in the story one tells to explain the results, therefore reinforcing existing social biases. 

Race was among those factors. In life insurance for instance, \cite[p.~34]{bouk} describes how, at the end of the 19th century, insurers charged the same premium to everyone but paid claims differentially according to skin colour (see also \cite{heen2009ending}). Several states then passed anti-discrimination laws. Thus, in the summer of 1884, the state of Massachusetts enacted a law prohibiting ``{\em any distinction or discrimination between white persons and colored persons wholly or partially of African descent, as to the premiums or rates charged for policies upon the lives of such persons}" (quoted in \cite[p.~68]{Wiggins}). To counter the law, Frederick L. Hoffman, supported by Prudential Life Insurance, published a book in 1896 demonstrating the higher mortality of black Americans (\cite[p.~49–52]{bouk}; \cite[p.~377]{heen2009ending}). Insuring them at the same rate as Whites would be statistically inequitable, he argued; not insuring them was therefore the only way to comply with the law, which in fact made black Americans uninsurable.

Another example of a rating based on stereotypes comes from a reading of the training manuals for underwriting in the 1970s. It reveals a description of women as unreliable, unstable in their work, incapable of making financial decisions, and dependent on their male partners for their living (\cite[p.~174]{horan2021insurance}). These mindsets legitimated the use of the male/female parameter in underwriting.

Moreover, in the controversies surrounding classification, \cite[p.~170–71]{horan2021insurance}) shows that pricing parameters also evolve in response to regulatory, political or social constraints: ``{\em the categories insurance companies used to create risk classifications throughout the twentieth century reflected changing political trends and social values, and not simply objective realities}.” Alternative classifications can be equally effective, highlighting the room for manoeuvre, but also the arbitrariness of the decisions left to practitioners: ``{\em Insurers can rate risks in many different ways depending on the stories they tell about which characteristics are important and which are not} (...) {\em The fact that the selection of risk factors is subjective and contingent upon narratives of risk and responsibility has in the past played a far larger role than whether or not someone with a wood stove is charged higher premiums}'' (\cite[p.~135]{glenn2003postmodernism}).

While Glenn insists on the subjective selection of factors, one can argue that this selection, that aspires to scientific exactitude and objectivity, is actually oriented by what is recognized as true and/or acceptable in a given time and place.

For \cite{schauer2006profiles} two types of stereotypes should be distinguished. Some generalizations are completely unfounded: generalizations based on a person's astrological sign, for example, are pure prejudice. But others have a statistical basis, when the probability of having a character $y$ knowing $x$ is significantly different from the case where nothing is known. From this perspective, the use of the male/female parameter remains legitimate because it is statistically significant for estimating a probability of death or of a car accident. Any classification based on variables that are effectively correlated with the risk would then be legitimate.
\cite{works1977whatever} warns, however, against ``proxy variables", as opposed to ``true variables" of risk. The latter would be more difficult to obtain, and thus replaced by simple approximations - leaving again the door open to all kinds of biases in pricing and underwriting:
``{\em Although the core concern of the underwriter is the human characteristics of the risk,} cheap screening indicators are adopted as surrogates for solid information {\em about the attitudes and values of the prospective insured} (...) {\em The invitations to underwriters to} {\em introduce prejudgments and biases and to indulge amateur psychological stereotypes are apparent. Even generalized underwriting texts include occupational, ethnic, racial, geographic, and cultural characterizations certain to give offense if publicly stated}'' (\cite[p.~471]{works1977whatever}, emphasis added).
In the 1980s US class actions against the use of the gender parameter, this was the approach taken by the plaintiffs. Their main argument was that the observed correlation between the cost of motor insurance claims and the gender of the driver was due to the lower mileage of women; mileage is the causal variable, and therefore legitimate, and not gender, which is only a biased approximation of the latter (\cite{horan2021insurance}). The underlying hypothesis here is that there are "true variables" of risk which explain accidents in a causal manner, all the others being invalid. 
The problem with this type of argument is that the existence of a direct causality is very difficult to establish, and therefore it is more of a judgement, again socially constructed, than a real scientific proof: for example, when Hoffman at the end of the 19th century highlighted a correlation between life expectancy and skin colour, he deduced the existence of an innate causality linked to the black race that made it more risky, whereas others would have sought environmental and social causes explaining the greater mortality of Blacks (\cite[p.~ 377]{heen2009ending}). Hence while some causal relations might be objective physical properties, others might again be the latest version of a socially accepted causation.
Moreover, the use of proxy variables can also come from an intention to avoid regulation. For instance, once racial parameters became forbidden, insurers and other financial institutions started delineating areas according to their racial occupation.  Ethnicity can indeed be inferred with a fair degree of accuracy from the location of potential policyholders. A survey commissioned by the federal government in the 1960s thus revealed that many financial institutions refused to serve predominantly African-American geographic areas (\cite{austin1983insurance}; \cite{horan2021insurance}), and that this systematic practice of “red-lining” had led to a deterioration of services and infrastructure in certain cities. This discrimination seems to persist to this day: \cite{larson2017we}conducted an analysis by zip code for major insurance companies across the US, showing that the average premium is 10\% higher in auto liability for zip codes associated with minority populations. 

Besides, even if a causality has been recognized as valid, it remains open to political controversies. \cite[p.~795-6]{simon1988ideological}) for instance argues that causality or correlation ultimately do not matter when it comes to fighting blatant social discrimination: on this basis, the use of the discriminating parameter contributes to naturalizing the difference in (social) treatment and thus to anchoring the discriminatory reality. This happens for instance with credit-based insurance scoring, that has given way to intensive debates in the United States, starting in the 1990s up to this day. While credit-based scores properly predict insurance losses, the National Association of Insurance Commissioners (NAIC) wanted to better understand why that was the case. In her study, \cite{kiviat2019moral} shows how competing causal theories were proposed in the numerous hearings: on the one hand, insurers defended the idea that higher credit scores reflect careful behavior; on the other, public representatives demonstrated that the scores predicted the propensity to file a claim in case of an accident, but not the likelihood of having an accident. Hence the ``real” cause is the economic status of the insured, which made credit-based insurance rating an unfair practice:  policymakers were indeed ``{\em less willing to accept that consumers might deserve to pay more by virtue of their low earning power}” (\cite[p.~1146]{kiviat2019moral}). 

In this understanding, a biased model reinforces existing discrimination. In a classical statistical set-up, the solution is to prohibit the use of the variable. The list of such protected factors varies greatly between countries. Currently it includes, in Europe, religious beliefs, sexual orientation, trade union involvement, ethnicity, medical status, criminal convictions and offences, biometric data, genetic information and, most recently, gender.

Interestingly, in the 2011 decision that prohibits the use of gender in insurance prices, the judge further distinguishes between two types of causal variables, pointing to what could be considered a fair classification, once non-significant and non-causal variables are eliminated: ``{\em Like race and ethnic origin, sex is also an inseparable characteristic of the insured person over which} he or she has no influence" (\cite{curia}, emphasis added). This distinction refers to what is known in the literature as ``{\em brute and option luck}” (\cite{dworkin1981equality}): some hazards are linked to personal choices (option luck) and are thus voluntary; others are caused by elements over which the individual has no control (brute luck). The judge thus implies that while the former can be used for pricing, the latter must be taken care of by the community (and therefore protected and eliminated from pricing).

This distinction is however even more difficult to establish than the causal inference. In the case of natural catastrophes for instance, some have argued that to live in a risky area is a choice, hence the pricing according to geography was legitimate; others however insist that some less well-off populations have no other choice but to live in cheaper yet riskier areas (\cite{charpentier2022assuranceGB}).

Another famous example comes from the incidence of smoking on lung cancers, and the potential use of smoking status in mortality tables. It did not happen before the 1980’s (\cite{society1983report}; \cite{miller1983life}; \cite{benjamin_michaelson_1988}), because the factor remained controversial for a long time, as \cite{fisher1958cancer} had warned against confusing correlation with causation. For him, the studies all showed the existence of a correlation with lung cancer, but did not prove that smoking was its cause: ``{\em it would equally be possible to infer on exactly similar grounds that inhaling cigarette smoke was a practice of considerable prophylactic value in preventing the disease, for the practice of inhaling is rarer among patients with cancer of the lung than with others}" (\cite{fisher1958cancer}). In his papers, he strives to show that a genetic configuration is at the origin of the surplus of cancers in the smoking population, that also inclines them to smoke. From the perspective of this paper, the debate around smoking proposed by Fisher highlights two types of potential bias: the fact that the smoker~/~non-smoker factor would not be a causal factor; the fact that if the cause is genetic, then it falls into the category of variables over which the individual has no control and should therefore be banned from the tariffs for reasons of fairness (regardless of the fact that the processing of genetic data is forbidden).

The critics of classical classifications are thus distributed along three types of biases: 

Type 1 biases are linked to classes that do not reflect the reality of the risk, either by mistake or due to pure prejudice. In statistical terms, the model is bluntly wrong. In social ones, it comes to perpetuate existing discriminations in the form of what \cite[p.~694]{barocas2016big} call disparate treatment:  ``{\em Disparate treatment comprises two different strains of discrimination: (1) formal disparate treatment of similarly situated people and (2) intent to discriminate}".

Type 2 biases are linked to classes that reflect a proven statistical reality (a correlation with risk), but the variables are known to be non-causal. This makes them suspect of bias and arbitrary choice, in the same way as type 1 are. This is the case for instance with redlining, where the variable is used as a proxy of race to discriminate against minorities.

Type 3 biases are linked to classes that reflect a statistical reality, that is taken to be causal. However, for some reason, the classification is deemed unacceptable on other grounds. Although utterly correct, it is intrinsically harmful because it reproduces and anchors in reality a situation that must be fought against. Type 3 biases also include cases where the causal variable does not capture an existing social discrimination, but its use for pricing would create one. Differentiating risks based on genetic data in health insurance would be such an example. Type 3 biases thus also establish a distinction between causal elements that are voluntary or non-voluntary, the pricing based on the latter being considered unfair.

Interestingly, the distinction between type 2 and 3 biases is as difficult as proving causality. The distinction thus comes from whether the explanation provided by the model makes (causal) sense in a given time and place. But this further triggers questions, such as - when is a correlation admitted as a cause, and when is it not? And, once recognized as a cause, when is it perceived as acceptable, and why? Trying to answer these questions would go far beyond the scope of this paper. However, we will try to show in the next section how these become crucial with machine learning.

\section{From Classification to Machine Learning}

From the 2000s onwards, big data and new algorithms have implied a displacement of tasks between humans and machines. Measuring the impact on insurance is however maybe more difficult than in other areas. On the one hand, like any other organization, insurers are pushed to change their practices to incorporate the new sources of data that have become available, the increased computing capacity, and the new algorithms. This has obvious consequences on the risk classification process.  On the other hand, the new techniques often appear to be the continuation of this almost age-old segmentation practice (\cite{swedloff2014risk}). From this perspective, we will show below that the new algorithms are only renewing existing questionings on discrimination and biases.

\subsection{The Datafication of Risk}

According to \cite{ewald2012assurance}, the new era means that the quantification of hazards, and the way to protect society against them, is changing: we would be moving from risk, where hazard were dealt with by classifying and mutualizing hazard into homogeneous classes, to ``data.”   The ``datafication” of risk (\cite{mayer2013big}) then means that hazards are not calculated via the classification process, but by extensive use of individual data. However, some studies show that, to date at least, pricing models have not changed significantly, nor have new products emerged beyond pilot experiments (\cite{barry2020personalization}; \cite{franccois2022revolution}). The study below is therefore more of an analysis of what the new models make possible, even if the change in insurance has not (yet?) been observed in practice, leading \cite{Meyers_Van_Hoyweghen_2018} to speak of ``not-yet markets.”

The change appears first as conceptual. For instance, in her opinion concerning the prohibition of the use of gender in insurance pricing, the European judge invokes the imprecision of statistics, when what is at stake, in her view, is the individual risk: ``{\em The life expectancy of insured persons} (...) {\em is strongly influenced by economic and social conditions as well as} by the habits of each individual {\em(for example, the kind and extent of the professional activity carried out, the family and social environment, eating habits, consumption of stimulants and/or drugs, leisure activities and sporting activities}" (\cite{curia}, emphasis added).

In other words, individual behaviour rather than statistics, which are necessarily aggregated, determine (individual) risk. The judge seems to encourage pricing based on the behavioural data recently made available by sensors. Such is indeed the conclusion of researchers:  
``{\em The gender directive gave telematics insurance products an added push. Since insurers are no longer able to differentiate by gender in their rating - and previously gender had quite an influence on the premium charged - telematics offers the opportunity to see how someone really drives. Hence, you get products now that are designed to reward better drivers, regardless of their gender. The likelihood is that, certainly at the younger end, more of the better drivers will be female}" (Munich Re, quoted in \cite{meyers2018enacting}).
The judge is thus in line with the trend that associates risk with lifestyle, and no longer sees it as determined based on statistical classes (\cite{rebert2015right}). This same trend appears in the US Prohibit Auto Insurance Discrimination (PAID) Act, which proposes that any parameter not directly related to driving should be prohibited in the pricing of automobile risk (\cite{metz2020sen}). The utopic claim that today's algorithms would be capable of personalizing decisions, whereas their ancestors were only grossly working on averages (\cite{lury2019algorithmic}; \cite{moor2018price}) is thus in the process of being transposed to insurance with, paradoxically, specific issues for fairness.

This conceptual shift is driven by the emergence of big data. In sharp contrast with the era of questionnaires, today data are obtained via sensors, connected objects or coming from online actions; they are thus natively digital - all sources that do not require human intervention. Moreover, these data are more often than not behavioural, and almost continuous data: for sensors for instance, telematics continuously collect the position, speed and acceleration of the vehicle (\cite{barry2020personalization}); connected bracelets or ``wearables” measure the biometric data of their wearers (\cite{lupton2014self}, \cite{lupton2016diverse}).

The second major change is that the computational capabilities of computers are now far greater than in the previous generation. Thus when \cite{de1984rate} referred to the possibility of refining segmentation, they relied on the idea that ``with the help of computers {\em it has become possible to make thorough risk analyses, and consequently to arrive at further premium differentiation}” (\cite[p.~155]{de1984rate}, emphasis added): it is the very existence of computers which, in the 1980s, changed the situation compared to an earlier era when calculations were practically manual (\cite{Barry2020InsuranceBD}; \cite{grier2013computers}). Today, the computing capacity that allows the processing of much larger databases is changing the environment of data analysis.

Finally, machine learning allows for the automation of (some?) of the tasks, particularly that of choosing significant variables, which increases the number of variables that can be considered. Models thus become more complex, without necessarily changing their nature. A conceptual leap occurs, however, with deep learning algorithms (taken here as a type of machine learning). \cite[p.~436]{lecun2015deep} indeed characterize deep learning by its ability to infer potential relationships between variables, where these were previously imposed on the data by the analyst: "the key aspect of deep learning is that these layers of features are not designed by human engineers: they are learned from data using a general-purpose learning procedure.”

Put in perspective with the previous section, machine learning seems to remove type 1 and 2 biases that resulted from the actuary's prejudices and stereotypes in his choice and coding of variables. The recent access to behavioural data also seems to meet the need to distinguish between variables describing conscious choices of the insured (his behaviour) and those relating to intrinsic characteristics over which he has no control, hence lifting part of the difficulties linked to type 3 biases and brute versus option luck. Would big data and machine learning thus allow to lift the discriminations associated with classification? We’ll show below that nothing is less certain, even if the claim that they could is recurringly heard.

\subsection{The biases of machine learning in insurance}

The biases delineated in the first section were linked respectively to social perceptions of the world that influenced the statistician, to correlations that were not causation, and to causal relations that should be corrected rather than reproduced in the models. Interestingly, each of these biases find a new expression and answer with machine learning. 

\subsection{Risks’ ``Ground Truth” }

``Ground truth” is an important and interesting concept of machine learning. At first hand, the meaning seems clear – it is the physical reality that the algorithm comes to predict. Wikipedia’s definition is ``{\em the knowledge of the truth concerning a specific question. It is the ideal expected result}.” It refers however to the ``ground truth” as seen by the algorithm, that is as mediated by the data taken to describe the physical reality at stake. But to what extent is this representation accurate? To what extent does ``ground truth” actually reflect reality?
Insurance companies are increasingly relying on data from external sources for modelling their risks: most frequently, these are online actions, be they invoices, transactions, emails, photos, click streams, logs, search queries, medical records, etc. (\cite{charpentier2022assuranceGB}). In the somewhat mythological utopia of big data, this data would finally have become exhaustive, allowing a richer (full?) account of reality: without the compromise of sampling, without the constraints of volume (\cite{mayer2013big}) and without the reduction of reality due to quantification problems (\cite{desrosieres2008argument}). 

One key issue however is that these data are observational, leading -paradoxically for the myth of exhaustivity, to multiple biases (\cite{Rosenbaum_2018}; \cite{charpentier2022assuranceGB}). The distinction between observational and experience data was first established by Ronald Fisher in 1935 (\cite{fisher1937design}) when devising randomized experiments, to solve type 2 biases and separate between causal and non-causal relationships. The method consists in setting up two similar (if not identical) groups with which to measure the effect of a variable (a treatment in medicine, for example). The proximity is measured by a series of co-variables (age, gender, economic status, etc), that all might affect the result and are hence ``under control.” One of the groups, known as the control group, is indeed not subjected to the treatment, allowing to observe the effect of the treatment on the second group, all other things being equal (\cite{Leigh}; \cite{Rosenbaum_2018}). In the smoking controversy, Fisher thus claimed that without comparing identical populations, the observation of a correlation between smoking and cancer proved nothing. 

Resulting, in most cases, from the behaviour of people independently of any ad-hoc experiment, big data are observational hence cannot, without further qualification, prove causation. An example of such a bias  in data science is the one given by \cite{caruana2015intelligible}: researchers in the 1990s trained a deep neural network to classify patients as low or high risk for pneumonia, in order to limit hospital admissions. The model was extremely accurate on the training data, but the results were counter-intuitive on asthma patients, who were classified as low risk. A closer examination showed that asthma patients were in practice treated much more quickly, precisely because of their very high mortality risk in case of pneumonia. Their low risk was therefore the result of differential treatment (the two populations were not identical) which the algorithm did not take into account. This example shows that one cannot fully make sense of the data without knowing the stories behind them, that is the process that led to their collection. For \cite[p.~206]{pearl2018book}, ``{\em we must look beyond the data to the data generating process.}” 

Retroaction bias also results from the generating process that escapes observational data: in health insurance, the Affordable Care Act introduced a penalization for hospitals that went past a threshold in readmissions within 30 days. As a result, hospitals started monitoring the indicator, keeping it below the threshold by registering readmissions under another code, thus modifying both variables without any obvious real change (\cite{poku2016campbell}). This would lead to models that cannot properly learn real life processes because the underlying causes of the change in pattern remain unknown.

Besides, big data magnify type 1 sample biases of the classical statistics era, such as Hoffman’s choosing specific or partial populations to compare mortality rates between Blacks and Whites (\cite{bouk}). With big data, the filtering is no longer the responsibility of the statistician who builds his database, yet it exists all the same (\cite{boyd2012critical}). For \cite{barocas2016big}, we have moved from disparate treatment, that is Hoffman consciously building ad hoc data to prove a point, to disparate impact. This happens non intentionally when populations at the margins of the formal economy and/or online activities are under-represented in the data used to train the algorithms. The non-intentionality and potential transparence of the problem to the expert aggravates the discrimination issue, the more so as models have become more complex.

Another known big data sample bias is related to self-selection (\cite{charpentier2022assuranceGB}). This situation is increasingly found in European public administrations’ data bases for instance, where until very recently data were stored automatically. With the General Data Protection Regulation (GDPR) - whose main aim is the protection of personal data, it is now possible for those who request it to have their data deleted. This concept of opting-out can strongly bias the data retained. In all these examples, the actuary, who has not himself constructed the databases to which he now has access, sometimes has difficulty understanding their limitations.

\subsection{Interpretability vs. Accuracy: the Opaque Effectiveness of New Algorithms}

The often decried opacity of the new algorithms is usually weighed in the debates against their increased accuracy compared to classical models (\cite{breiman2001statistical}). In 2017, during one of the first debates at the NeurIPS conference, it was pointed out that ``{\em if we wish to make AI systems deployed on self-driving cars safe, straightforward black-box models will not suffice, as we need methods of understanding their rare but costly mistakes.}” At the conference, Yann LeCun claimed that, when presented with two models (one perfectly interpretable and 90\% accurate, the other a black-box with a higher accuracy of 99\%), people always chose the more accurate model. He concludes that ``{\em people don't really care about interpretability but just want some sort of reassurance from the working model}" (\cite{caruana93great}). Interpretability is not important once one is convinced that the model works well under the conditions in which it is supposed to work. This claim completely displaces the understanding of type 2 biases (the attempt to separate correlation from causation) in the big data era.

For \cite[p.~3]{napoletani2011agnostic}, this is a new scientific paradigm, opening the way to a new kind of science that has become ``agnostic" to causes. As such, in a now famous article \cite{anderson2008end} speaks of the “end of theories” to characterize this new approach, that leads to the renunciation of the need to highlight causal links between variables (and therefore to explain the phenomenon) of the analysis: ``{\em
Scientists are trained to recognize that correlation is not causation, that no conclusions should be drawn simply on the basis of correlation between $X$ and $Y$ (it could just be a coincidence). Instead, you must understand the underlying mechanisms that connect the two. Once you have a model, you can connect the data sets with confidence. Data without a model is just noise. But} faced with massive data, this approach to science - hypothesize, model, test - is becoming obsolete'' (\cite{anderson2008end}, emphasis added).

In the perspective of this paper, this would mean giving up the explicitation of the relationships between variables, either causal or correlated: the machine produces a score, sufficiently precise to justify giving up an interpretation by the input data. There are a few (rare) examples of this black-box approach in insurance. For example, image recognition algorithms can infer risk factors. \cite{kita2019google} thus propose to predict the frequency of car accidents from satellite images of the driver's place of residence; or \cite{CapGemini2020} calculates a health score based on a photo portrait of the person. 

The efficacity of black-box algorithms cannot however lift the discrimination issues, to the contrary. Paradoxically, while in the previous era, the stories told by statisticians were incriminated as leading to socially constructed biases, it is now the impossibility of telling (causal) stories that poses a problem. As \cite[p.~1152]{kiviat2019moral} puts it ``{\em what gets wiped away along with storytelling is an ability to appreciate how bad luck or the inequities of history can set events in motion and cause people to show up in the data in particular ways}.” 

\subsection{Correcting social discrimination: protecting data in a big data environment}

Moreover, while the machine learning process is supposed to be agnostic to social constructions and biases, we now know that, on the contrary, prejudices, stereotypes and other discriminations are found in the data themselves, and therefore well upstream of the statisticians' judgement: beyond the sample bias mentioned above, it is really the nature of the data that is at issue (\cite{Caliskan183}). 
Besides, whereas in classical models one could hope to correct biases by prohibiting the use of so-called protected variables, the correlation of these variables with other, facially neutral variables in big data makes this ``protection" illusory: ``{\em thus, a data mining model with a large number of variables will determine the extent to which membership in a protected class is relevant to the sought-after trait} whether or not that information is an input" (\cite{barocas2016big}, emphasis added).

For \cite{prince2019proxy} the proxy discrimination, already mentioned for classical models, is magnified by the new algorithms. Whereas it was intentional in the past (since a human decision presided over the choice of variables), proxy discrimination becomes unintentional. This phenomenon is unavoidable, especially when a variable directly related to the phenomenon (a causal variable) is absent from the data: 
To illustrate, an AI deprived of information about a person's genetic test results or obvious proxies for this information (like family history) will use other information-ranging from TV viewing habits to spending habits to geolocational data-to proxy for the directly predictive information contained within the genetic test results (\cite[p.~1274]{prince2019proxy}).

In this specific example, the algorithm might create a health risk factor based on the viewing of television programs! To combat this phenomenon, \cite{williams2018algorithms} show that the collection and use of protected variables should not be prohibited, but rather used as a means of monitoring non-discrimination instead. But this is not easily done either, as will be shown below.

\section{Insurance Fairness in the Age of Machine Learning: Collective or Individual?}

According to \cite{thiery2006fairness}, lawyers and actuaries have fundamentally different conceptions of equity. Legally, the right to equal treatment is granted to a person as an individual. But this view is fundamentally opposed to actuarial fairness, which historically builds on an analysis of risks and calculation of premiums in collective terms (\cite{ewald2012assurance}). For instance, in the 1983 Norris decision on a class action against the use of gender for retirement benefits, the judge maintained that a statistically valid classification (whether causal or correlational) does not make it a legitimate classification. In fact, no classification can be, since ``{\em even a true generalization about class cannot justify class-based treatment. An individual woman may not be paid lower monthly benefits simply because women as a class live longer than men}” (quoted in \cite[p.~187]{horan2021insurance}). For the individual on whom it is imposed, classification always results from a ``statistical bias” (\cite{binns2018fairness}), that is an arbitrary inference from the group to the individual. 

For \cite{simon1988ideological} and \cite{horan2021insurance}, the adoption of this individual point of view by the judge in the Norris case has contributed to reinforcing the erasure of the principle of solidarity which is at the heart of insurance practice (\cite{lehtonen2011forms}). Once the existence of an individual risk has been accepted, to be approached through classification and, later, through learning algorithms, pricing becomes a mathematical exercise in optimizing and minimizing the insurer’s variance. Big data algorithmic ``personalization” is translated into risk ``individualization” in insurance (\cite{barry2020personalization}). Thus the distinction between collective-insurance versus legal-individual kinds of fairness tends to disappear (\cite{Barry2020bigdata}), giving to the {\em statistical discrimination} outlined by the judge in Norris a renewed importance. 

Indeed, even if actuaries have not fundamentally changed their practice, the notion of insurance fairness is changing. \cite{Meyers_Van_Hoyweghen_2018} thus show on a telematic product that risk is no longer presented as a pooled uncertainty, but as an individual choice. Individual behaviour should determine the premium, not aggregate demographics. Fairness in this case means adjusting the premium to individual behaviour, so that everyone pays according to ``their" risk (\cite{meyers2018enacting}). For \cite[p.~ 24]{fourcade2017categories}, algorithmic scores rely on moral economy of deserve, where each is responsible for his acts and their consequences.

This individualization, if it takes place, triggers its own fairness and discrimination issues. First, since it would lead to more disparate pricing, individuals found as ``the riskiest” by the algorithm are likely to get unaffordable premiums, hence excluding them from the insured community (\cite{charpentierAM1}). Besides, research in algorithmic fairness, that is emerging as a new discipline \cite{kusner2020long}, shows that the tension between individual and collective viewpoints, at the heart of current questionings on the individualization of risk, interestingly finds a new forum in this literature. 

Aggregate indicators of the fairness of an algorithm might come from its (mathematical) accuracy. It is generally measured by means of a confusion matrix, which allows to observe errors by type - false negatives and false positives. But simultaneously minimizing these errors is not possible, or even desirable, for several reasons. Indeed, false positives and false negatives are not comparable from an ethical point of view: the conviction of an innocent person does not have the same ``value" as the release of a guilty person. Thus, depending on the context, it will be necessary to choose to minimize one or the other form of error.

Things become even more complicated when protected variables are taken into account \cite{pessach2020algorithmic}. The researchers then distinguish between collective and individual fairness indicators. Collective ones aim to ensure parity between protected and non-protected groups. One can thus check that the (exact or positive) prediction frequencies, or that the false positive and false negative rates, calculated separately for each group are close. This was not the case, for example, for the Correctional Offender Management Profiling for Alternative Sanctions algorithm (COMPAS), which had a much higher false positive rate (falsely classified as recidivists) for Blacks, and a higher false negative rate (falsely classified as non-recidivists) for Whites, with equal accuracies on both groups (\cite{kleinberg2016inherent}). Individual indicators, on the other hand, aim to ensure that similar individuals obtain a similar score. \cite{kusner2020long} thus define ``{\em counterfactual}'' equity, which consists of comparing the scores of two identical observations in which only the protected variable takes a different value. All authors agree that these different indicators cannot be optimized simultaneously, leading to necessary context-dependent trade-offs (\cite{kleinberg2016inherent}; \cite{pessach2020algorithmic}).

The ban on the male/female parameter by the European directive exemplifies these dilemmas in insurance: either the variable is ignored, but then if a statistical difference exists it will be captured through other variables, that are correlated with the banned parameter. Consequently, the average for men and women will remain different. Or, on the contrary, this variable is used to maintain identical averages, but then, all other things being equal, the rates will vary with the sex of the person. It will never be possible to maintain parity between the groups and ensure counterfactual equity. For \cite[p.~148]{charpentier2022assuranceGB}, prohibiting the use of the protected variable is counterproductive because ``{\em in most realistic cases, not only does the removal of the sensitive variable not make the regression models fair, but on the contrary, such a strategy is likely to amplify the discrimination}". 

\section{Conclusion}

Insurance fairness is a dynamic concept, which depends on historical, cultural and technical contexts. At the height of the industrial era, the veil of ignorance on individual hazards and the idea of the equal fate of all in front of adversity were used to justify very broad and solidary coverage. This conception of fairness was criticized by the most conservatives since it was seen as an incentive to license. During the 20th century, with the growing capacities of data gathering and calculation, segmented models were adopted, which based insurance on the classification of risks into homogeneous groups of people. From the 1980s onwards, controversies arose over the use of this or that variable, that frame the current criticism of machine learning’s biases and discriminations. The examination of this history allowed us to identify in traditional classification practices a few main families of bias, that could then be tracked in machine-learning algorithms. 

The first claim is that when the parameters have no connection with the phenomenon to be studied, their (manual) choice only reflect the prejudices of the statistician. This is the case of skin colour in the United States in life insurance products at the end of the 19th century. This type of bias disappears in principle with big data: being natively digital, they bypass the quantification work of the previous period. However, over the last twenty years, researchers have shown that data absorb social discrimination and prejudices. Blind use of machine learning would then lead to the reproduction of these biases in algorithmic decisions.

It was soon found that discarding statistically irrelevant variables is not enough. In the 1960s and 1980s, the use of correlated but not causal variables, that is magnified with the new algorithms, became the main source of complaint: the male/female parameters, the credit score or the smoker/non-smoker criterion have all provoked controversies of this kind, some of which continue to this day. The requirement of proven causality, already inoperable in the statistical era, has been totally abandoned in machine-learning algorithms. Some indeed say that big data marks the advent of a new episteme, where finding correlations and patterns in the input data, without even making these links explicit, would be the core of science. But this displaces the criticism towards the algorithms’ opacity, that despite greater precision create their own implicit biases.

Moreover, some causal variables reflect hazard that were not chosen by the ones that endure them. In this case, insurance is seen as a way not to reflect the risk but, by eliminating the variable from the models, to have it borne by the entire insured population: the use of genetic data in health insurance, for example, is prohibited today in most countries. But the solution of eliminating protected variables, effective in traditional models, is much more difficult to implement with big data and machine learning, respectively because protected variables are captured via their correlation with others, and the opacity of the algorithms makes it more complex to highlight these discriminations. 

More fundamentally, a legalistic critique opposed fairness as an individual right to the collective approach taken by insurance. In this strand of thought, the necessarily arbitrary reduction of an individual to the data of a class was seen as a statistical bias. Big and behavioral data, that sign the advent of personalization and the potential individualization of risk, supposedly solve this statistical bias by promoting in insurance the fairness of deserve, where each pay for the risks he chooses to take. But here again, the promise is not kept: researchers in algorithmic fairness indeed highlight the impossibility of optimizing algorithms on multiple criteria, none of which can be a priori preferred to another. In the insurance context, individual fairness also threatens to lead to increasingly differentiated rates, therefore making insurance unaffordable for those classified as very risky.

This history also interestingly shows that once segmentation is adopted as a principle, the legitimacy of an insurance model relies on its being both “causal,” in the sense of explaining risk, and fair. Yet both causality and fairness are strongly contingent upon the narratives seen as valid in a given time and place. Besides, current state of the art algorithms cannot produce models that satisfactorily provide (causal) explanations. Should insurance then stick to the good old pricing tables, for which all the parameters are explicit, known in advance and therefore open to challenge? This does not warrant fairness, but contestability: just as any scientific theory must be falsifiable, a pricing system should be transparent in order to be contestable.

\bibliographystyle{apalike}
\bibliography{sample}

\begin{thebibliography}{}

\bibitem[Anderson, 2008]{anderson2008end}
Anderson, C. (2008).
\newblock The end of theory: The data deluge makes the scientific method
  obsolete.
\newblock {\em Wired magazine}, 16(7):16--07.

\bibitem[Austin, 1983]{austin1983insurance}
Austin, R. (1983).
\newblock The insurance classification controversy.
\newblock {\em University of Pennsylvania Law Review}, 131(3):517--583.

\bibitem[Baker and Simon, 2002]{baker2002embracing}
Baker, T. and Simon, J. (2002).
\newblock {\em Embracing Risk: The Changing Culture of Insurance and
  Responsibility. Chicago: Univ}.
\newblock Chicago Press.

\bibitem[Barocas and Selbst, 2016]{barocas2016big}
Barocas, S. and Selbst, A.~D. (2016).
\newblock Big data's disparate impact.
\newblock {\em California Law Review}, 104:671--732.

\bibitem[Barry, 2020]{Barry2020InsuranceBD}
Barry, L. (2020).
\newblock Insurance, big data and changing conceptions of fairness.
\newblock {\em European Journal of Sociology}, 61:159 -- 184.

\bibitem[Barry and Charpentier, 2020]{barry2020personalization}
Barry, L. and Charpentier, A. (2020).
\newblock Personalization as a promise: Can big data change the practice of
  insurance?
\newblock {\em Big Data \& Society}, 7(1):2053951720935143.

\bibitem[Barry and {Charpentier}, 2020]{Barry2020bigdata}
Barry, L. and {Charpentier}, A. (2020).
\newblock {Personalization as a promise: Can Big Data change the practice of
  insurance? }.
\newblock {\em Big Data \& Society}.

\bibitem[Benjamin and Michaelson, 1988]{benjamin_michaelson_1988}
Benjamin, B. and Michaelson, R. (1988).
\newblock Mortality differences between smokers and non-smokers.
\newblock {\em Journal of the Institute of Actuaries}, 115(3):519–525.

\bibitem[Binns, 2018]{binns2018fairness}
Binns, R. (2018).
\newblock Fairness in machine learning: Lessons from political philosophy.
\newblock In {\em Conference on Fairness, Accountability and Transparency},
  pages 149--159. PMLR.

\bibitem[Bouk, 2015]{bouk}
Bouk, D. (2015).
\newblock {\em How Our Days Became Numbered: Risk and the Rise of the
  Statistical Individual}.
\newblock The University of Chicago Press.

\bibitem[Boyd and Crawford, 2012]{boyd2012critical}
Boyd, D. and Crawford, K. (2012).
\newblock Critical questions for big data: Provocations for a cultural,
  technological, and scholarly phenomenon.
\newblock {\em Information, communication \& society}, 15(5):662--679.

\bibitem[Breiman, 2001]{breiman2001statistical}
Breiman, L. (2001).
\newblock Statistical modeling: The two cultures (with comments and a rejoinder
  by the author).
\newblock {\em Statistical science}, 16(3):199--231.

\bibitem[Caliskan et~al., 2017]{Caliskan183}
Caliskan, A., Bryson, J.~J., and Narayanan, A. (2017).
\newblock Semantics derived automatically from language corpora contain
  human-like biases.
\newblock {\em Science}, 356(6334):183--186.

\bibitem[Caruana and LeCun, 2017]{caruana93great}
Caruana, R. and LeCun, Y. (2017).
\newblock The great ai debate “interpretable ml symposium”.

\bibitem[Caruana et~al., 2015]{caruana2015intelligible}
Caruana, R., Lou, Y., Gehrke, J., Koch, P., Sturm, M., and Elhadad, N. (2015).
\newblock Intelligible models for healthcare: Predicting pneumonia risk and
  hospital 30-day readmission.
\newblock In {\em Proceedings of the 21th ACM SIGKDD international conference
  on knowledge discovery and data mining}, pages 1721--1730.

\bibitem[Charpentier, 2022]{charpentier2022assuranceGB}
Charpentier, A. (2022).
\newblock {\em Insurance: biases, discrimination and fairness}.
\newblock Institut Louis Bachelier.

\bibitem[Charpentier et~al., 2020]{charpentierAM1}
Charpentier, A., Barry, L., and Gallic, E. (2020).
\newblock Quel avenir pour les probabilités prédictives en assurance ?
\newblock {\em Annales des Mines}, (2020).

\bibitem[Charpentier and Denuit, 2005]{charpentierMASSNV2}
Charpentier, A. and Denuit, M. (2005).
\newblock {\em Mathématiques de l’assurance non-vie - Principes fondamentaux
  de théorie du risque (Tome 2)}.
\newblock Economica.

\bibitem[Charpentier et~al., 2014]{charpentier103}
Charpentier, A., Denuit, M., and \'Elie, R. (2014).
\newblock Segmentation et mutualisation, les deux faces d’une m\^eme pi\`ece.
\newblock {\em Risques}, (103).

\bibitem[CURIA, 2011]{curia}
CURIA (2011).
\newblock Association belge des consommateurs test-achats asbl, vann van vugt,
  charles basselier v conseil des ministres.
\newblock {\em eur-lex.europa.eu}, C-236/09 ECJ.

\bibitem[De~Pril and Dhaene, 1996]{de1996segmentering}
De~Pril, N. and Dhaene, J. (1996).
\newblock Segmentering in verzekeringen.
\newblock {\em DTEW Research Report 9648}, pages 1--56.

\bibitem[De~Wit and Van~Eeghen, 1984]{de1984rate}
De~Wit, G. and Van~Eeghen, J. (1984).
\newblock Rate making and society's sense of fairness.
\newblock {\em ASTIN Bulletin: The Journal of the IAA}, 14(2):151--163.

\bibitem[Desrosi{\`e}res, 2008]{desrosieres2008argument}
Desrosi{\`e}res, A. (2008).
\newblock {\em L’argument statistique I: Pour une sociologie historique de la
  quantification}.
\newblock Presses de l’École des Mines.

\bibitem[Dworkin, 1981]{dworkin1981equality}
Dworkin, R. (1981).
\newblock What is equality? part 1: Equality of welfare.
\newblock {\em Philosophy \& public affairs}, pages 185--246.

\bibitem[Ewald, 2012]{ewald2012assurance}
Ewald, F. (2012).
\newblock Assurance, prevention, prediction... dans l’univers du big data.
\newblock {\em Paris: Institut Montparnasse}.

\bibitem[Fisher, 1958]{fisher1958cancer}
Fisher, R. (1958).
\newblock Cancer and smoking.
\newblock {\em Nature}, 182(4635):596--596.

\bibitem[Fisher et~al., 1937]{fisher1937design}
Fisher, R.~A. et~al. (1937).
\newblock The design of experiments.
\newblock {\em The design of experiments.}, (2nd Ed).

\bibitem[Fourcade and Healy, 2017]{fourcade2017categories}
Fourcade, M. and Healy, K. (2017).
\newblock Categories all the way down.
\newblock {\em Historical Social Research/Historische Sozialforschung}, pages
  286--296.

\bibitem[Fran{\c{c}}ois and Voldoire, 2022]{franccois2022revolution}
Fran{\c{c}}ois, P. and Voldoire, T. (2022).
\newblock The revolution that did not happen. telematics and car insurance in
  the 2010s.
\newblock Technical report, Working Paper 26. Paris: Chaire PARI.

\bibitem[Frezal and Barry, 2019]{Frezal_Barry_2019}
Frezal, S. and Barry, L. (2019).
\newblock Fairness in uncertainty: Some limits and misinterpretations of
  actuarial fairness.
\newblock {\em Journal of Business Ethics}.

\bibitem[Glenn, 2000]{glenn2000shifting}
Glenn, B.~J. (2000).
\newblock The shifting rhetoric of insurance denial.
\newblock {\em Law and Society Review}, pages 779--808.

\bibitem[Glenn, 2003]{glenn2003postmodernism}
Glenn, B.~J. (2003).
\newblock Postmodernism: the basis of insurance.
\newblock {\em Risk Management and Insurance Review}, 6(2):131--143.

\bibitem[Grier, 2013]{grier2013computers}
Grier, D.~A. (2013).
\newblock {\em When computers were human}.
\newblock Princeton University Press.

\bibitem[Heen, 2009]{heen2009ending}
Heen, M.~L. (2009).
\newblock Ending jim crow life insurance rates.
\newblock {\em northwestern Jounral of Law \& Social Policy}, 4:360.

\bibitem[Horan, 2021]{horan2021insurance}
Horan, C. (2021).
\newblock {\em Insurance Era: Risk, Governance, and the Privatization of
  Security in Postwar America}.
\newblock University of Chicago Press.

\bibitem[Kita and Kidzi{\'n}ski, 2019]{kita2019google}
Kita, K. and Kidzi{\'n}ski, {\L}. (2019).
\newblock Google street view image of a house predicts car accident risk of its
  resident.
\newblock {\em arXiv}, 1904.05270.

\bibitem[Kiviat, 2019]{kiviat2019moral}
Kiviat, B. (2019).
\newblock The moral limits of predictive practices: The case of credit-based
  insurance scores.
\newblock {\em American Sociological Review}, 84(6):1134--1158.

\bibitem[Kleinberg et~al., 2016]{kleinberg2016inherent}
Kleinberg, J., Mullainathan, S., and Raghavan, M. (2016).
\newblock Inherent trade-offs in the fair determination of risk scores.
\newblock {\em arXiv}, 1609.05807.

\bibitem[Kranzberg, 1986]{kranzberg1986technology}
Kranzberg, M. (1986).
\newblock Technology and history:" kranzberg's laws".
\newblock {\em Technology and culture}, 27(3):544--560.

\bibitem[Kusner and Loftus, 2020]{kusner2020long}
Kusner, M.~J. and Loftus, J.~R. (2020).
\newblock The long road to fairer algorithms.
\newblock {\em Nature}.

\bibitem[Larson et~al., 2017]{larson2017we}
Larson, J., Angwin, J., Kirchner, L., and Mattu, S. (2017).
\newblock How we examined racial discrimination in auto insurance prices.
\newblock {\em ProPublica, April}, 5.

\bibitem[LeCun et~al., 2015]{lecun2015deep}
LeCun, Y., Bengio, Y., and Hinton, G. (2015).
\newblock Deep learning.
\newblock {\em nature}, 521(7553):436--444.

\bibitem[Lehtonen and Liukko, 2011]{lehtonen2011forms}
Lehtonen, T.-K. and Liukko, J. (2011).
\newblock The forms and limits of insurance solidarity.
\newblock {\em Journal of Business Ethics}, 103(1):33--44.

\bibitem[Leigh, 2018]{Leigh}
Leigh, A. (2018).
\newblock {\em Randomistas: How Radical Researchers Are Changing Our World}.
\newblock Yale University Press.

\bibitem[Lupton, 2014]{lupton2014self}
Lupton, D. (2014).
\newblock Self-tracking modes: Reflexive self-monitoring and data practices.
\newblock {\em Available at SSRN 2483549}.

\bibitem[Lupton, 2016]{lupton2016diverse}
Lupton, D. (2016).
\newblock The diverse domains of quantified selves: self-tracking modes and
  dataveillance.
\newblock {\em Economy and Society}, 45(1):101--122.

\bibitem[Lury and Day, 2019]{lury2019algorithmic}
Lury, C. and Day, S. (2019).
\newblock Algorithmic personalization as a mode of individuation.
\newblock {\em Theory, Culture \& Society}, 36(2):17--37.

\bibitem[Mayer-Sch{\"o}nberger and Cukier, 2013]{mayer2013big}
Mayer-Sch{\"o}nberger, V. and Cukier, K. (2013).
\newblock {\em Big data: A revolution that will transform how we live, work,
  and think}.
\newblock Houghton Mifflin Harcourt.

\bibitem[Metz, 2020]{metz2020sen}
Metz, J. (2020).
\newblock Sen. booker’s paid act looks to eliminate discriminatory
  non-driving factors in auto insurance pricing.
\newblock {\em Forbes Advisor}, 5.

\bibitem[Meyers and Van~Hoyweghen, 2018a]{Meyers_Van_Hoyweghen_2018}
Meyers, G. and Van~Hoyweghen, I. (2018a).
\newblock Enacting actuarial fairness in insurance: From fair discrimination to
  behaviour-based fairness.
\newblock {\em Science as Culture}, 27(4):413–438.

\bibitem[Meyers and Van~Hoyweghen, 2018b]{meyers2018enacting}
Meyers, G. and Van~Hoyweghen, I. (2018b).
\newblock Enacting actuarial fairness in insurance: From fair discrimination to
  behaviour-based fairness.
\newblock {\em Science as Culture}, 27(4):413--438.

\bibitem[Miller and Gerstein, 1983]{miller1983life}
Miller, G. and Gerstein, D.~R. (1983).
\newblock The life expectancy of nonsmoking men and women.
\newblock {\em Public Health Reports}, 98(4):343.

\bibitem[Miller, 2009]{miller2009disparate}
Miller, M.~J. (2009).
\newblock Disparate impact and unfairly discriminatory insurance rates.
\newblock In {\em Casualty Actuarial Society E-Forum, Winter 2009}, page 276.
  Citeseer.

\bibitem[Moor and Lury, 2018]{moor2018price}
Moor, L. and Lury, C. (2018).
\newblock Price and the person: Markets, discrimination, and personhood.
\newblock {\em Journal of Cultural Economy}, 11(6):501--513.

\bibitem[Napoletani et~al., 2011]{napoletani2011agnostic}
Napoletani, D., Panza, M., and Struppa, D.~C. (2011).
\newblock Agnostic science. towards a philosophy of data analysis.
\newblock {\em Foundations of Science}, 16(1):1--20.

\bibitem[of~Actuaries, 1983]{society1983report}
of~Actuaries, S. (1983).
\newblock {\em Report of the Task Force on Smoker/Non-Smoker Mortality}.
\newblock Society of Actuaries.

\bibitem[Pearl and Mackenzie, 2018]{pearl2018book}
Pearl, J. and Mackenzie, D. (2018).
\newblock {\em The book of why: the new science of cause and effect}.
\newblock Basic books.

\bibitem[Pessach and Shmueli, 2020]{pessach2020algorithmic}
Pessach, D. and Shmueli, E. (2020).
\newblock Algorithmic fairness.
\newblock {\em arXiv}, 2001.09784.

\bibitem[Poku, 2016]{poku2016campbell}
Poku, M. (2016).
\newblock Campbell’s law: implications for health care.
\newblock {\em Journal of health services research \& policy}, 21(2):137--139.

\bibitem[Prince and Schwarcz, 2019]{prince2019proxy}
Prince, A.~E. and Schwarcz, D. (2019).
\newblock Proxy discrimination in the age of artificial intelligence and big
  data.
\newblock {\em Iowa Law Review}, 105:1257.

\bibitem[Rebert and Van~Hoyweghen, 2015]{rebert2015right}
Rebert, L. and Van~Hoyweghen, I. (2015).
\newblock The right to underwrite gender: The goods \& services directive and
  the politics of insurance pricing.
\newblock {\em Tijdschrift Voor Genderstudies}, 18(4):413--431.

\bibitem[Rosenbaum, 2018]{Rosenbaum_2018}
Rosenbaum, P. (2018).
\newblock {\em Observation and experiment}.
\newblock Harvard University Press.

\bibitem[Schauer, 2006]{schauer2006profiles}
Schauer, F. (2006).
\newblock {\em Profiles, probabilities, and stereotypes}.
\newblock Harvard University Press.

\bibitem[Shikhare, 2021]{CapGemini2020}
Shikhare, S. (2021).
\newblock Next generation ltc - life insurance underwriting using facial score
  model.
\newblock In {\em Insurance Data Science conference}.

\bibitem[Simon, 1988]{simon1988ideological}
Simon, J. (1988).
\newblock The ideological effects of actuarial practices.
\newblock {\em Law \& Society Review}, 22:771.

\bibitem[Swedloff, 2014]{swedloff2014risk}
Swedloff, R. (2014).
\newblock Risk classification's big data (r) evolution.
\newblock {\em Connecticut Insurance Law Journal}, 21:339.

\bibitem[Thiery and Van~Schoubroeck, 2006]{thiery2006fairness}
Thiery, Y. and Van~Schoubroeck, C. (2006).
\newblock Fairness and equality in insurance classification.
\newblock {\em The Geneva Papers on Risk and Insurance-Issues and Practice},
  31(2):190--211.

\bibitem[Walters, 1981]{walters1981risk}
Walters, M.~A. (1981).
\newblock Risk classification standards.
\newblock In {\em Proceedings of the Casualty Actuarial Society}, volume~68,
  pages 1--18.

\bibitem[Wiggins, 2013]{Wiggins}
Wiggins, B. (2013).
\newblock {\em Managing Risk, Managing Race: Racialized Actuarial Science in
  the United States, 1881–1948}.
\newblock University of Minnesota PhD thesis.

\bibitem[Williams et~al., 2018]{williams2018algorithms}
Williams, B.~A., Brooks, C.~F., and Shmargad, Y. (2018).
\newblock How algorithms discriminate based on data they lack: Challenges,
  solutions, and policy implications.
\newblock {\em Journal of Information Policy}, 8:78--115.

\bibitem[Works, 1977]{works1977whatever}
Works, R. (1977).
\newblock Whatever's fair-adequacy, equity, and the underwriting prerogative in
  property insurance markets.
\newblock {\em Nebraska Law Review}, 56:445.

\end{thebibliography}

\end{document}